\documentstyle[emulateapj,onecolfloat]{article}

\newcommand{\da}{d_A}

\newlength{\tskip}
\newlength{\colwidth}

\newcommand{\beq}{\begin{equation}}
\newcommand{\eeq}{\end{equation}}
\newcommand{\beqa}{\begin{eqnarray}}
\newcommand{\eeqa}{\end{eqnarray}}

\newcommand{\rad}{r}    

\begin{document}
\twocolumn[
\title{Anisotropy Studies of the Unresolved Far-infrared Background}
\author{Alexandre Amblard, Asantha Cooray}
\affil{
Center for Cosmology, University of California, Irvine, CA 92697}

\begin{abstract}
Dusty, starforming galaxies and active galactic nuclei that contribute
to the integrated background intensity at far-infrared wavelengths
trace the large-scale structure. Below the point source detection
limit, correlations in the large-scale structure lead to clustered
anisotropies in the unresolved component of the far-infrared
background (FIRB). The angular power spectrum of the FIRB anisotropies
could be measured in large-area surveys with the Spectral and
Photometric Imaging Receiver (SPIRE) on the upcoming Herschel
observatory.  To study statistical properties of these anisotropies,
the confusion from foreground Galactic dust emission needs to be
reduced even in the ``cleanest'' regions of the sky.  The
multi-frequency coverage of SPIRE allows the foreground dust to be
partly separated from the extragalactic background composed of dusty
starforming galaxies as well as faint normal galaxies.  The separation
improves for fields with sizes greater than a few hundred square
degrees and when combined with Planck data.  We show that an area of
about $\sim$ 400 degrees$^2$ observed for about 1000 hours with
Herschel-SPIRE and complemented by Planck provides maximal information
on the anisotropy power spectrum. We discuss the scientific studies
that can be done with measurements of the unresolved FIRB anisotropies
including a determination of the large scale bias and the small-scale
halo occupation distribution of FIRB sources with fluxes below the
point-source detection level.
\end{abstract}

]

\section{Introduction}
The total intensity of the extragalactic background light at far-IR
wavelengths is now established with absolute photometry (Puget et
al. 1996; Fixsen et al. 1998; Dwek et al. 1998), while deep surveys
with existing or previous instruments have resolved the cosmic far-IR
background (FIRB) to discreet sources at various fractions given the
wavelength (see reviews in Blain et al. 2002; 
Hauser \& Dwek 2001; Lagache et al. 2005).
Based on these results, the FIRB light is believed to be mostly due to
the thermal emission from interstellar dust in $z \sim 1$ to 3
galaxies with dust heated by ultraviolet radiation from stars and
active galactic nuclei. The far-IR source counts also include a
contribution from low-redshift spiral galaxies (Lagache et al. 2005).

Unfortunately even the deepest images of far-IR sky using instruments
on board the Herschel
observatory\footnote{http://herschel.esac.esa.int/} will be limited by
source confusion. For example, at 350 $\mu$m, at most 10\% of the
total background intensity will be resolved to individual sources
(e.g., Lagache et al. 2003).  To study the properties of the sources
that dominate the background light, we must consider the statistics of
the unresolved component.

In this respect, a useful statistic associated with the unresolved
background is the angular power spectrum of FIRB anisotropies (Haiman
\& Knox 2000; Knox et al. 2001; Scott \& White 1999; Negrello et al. 2007).
Unresolved far-IR background sources are expected to trace the
correlated large-scale structure and these correlations will be
reflected in the unresolved fluctuations. Based on previous models,
these fluctuations are expected to be at the level of
5\% to 10\% of the mean intensity at sub-degree angular scales (Haiman
\& Knox 1999). As discussed in Knox et al. (2001), the amplitude and
the shape of the FIRB anisotropy spectrum capture to some extent
informations on the number counts of sources and their redshift
distribution below the point source detection level, while a detailed
multi-frequency analysis of far-IR anisotropies can be performed to
establish the spectral energy distribution (SED) of dust characterized
by a mean temperature and a departure from the black-body spectrum.

Since the study of Knox et al. (2001), detailed phenomenological
models have been developed to describe the galaxy distribution in
large-scale structure through their connection to the underlying dark
matter halo distribution based on the halo model (see review in Cooray
\& Sheth 2002).  The halo model allows one to describe
the galaxy clustering power spectrum through the halo occupation
number or the number of galaxies in a dark matter halo as a function
of the halo mass. The occupation number description can be further
extended to account for the galaxy distribution in a given dark matter
halo mass as a function of the luminosity through what are now called
conditional luminosity functions (CLFs; Cooray 2005). With CLFs tuned
to reproduce the far-IR 350 $\mu$m luminosity functions of Lagache et
al. (2003), we extend the halo model to describe clustering at longer
wavelengths and study how clustering measurements of unresolved
fluctuations can be modeled, and informations extracted, through the
halo model (Cooray \& Sheth 2002).

Our study is mostly motivated by the possibility to study FIRB
anisotropies in near future with wide-field scan maps at 250 $\mu$m,
350 $\mu$m, and 500 $\mu$m from the Spectral and Photometric Imaging
Receiver (SPIRE; Griffin et al. 2006) aboard the Herschel observatory.
A challenge for anisotropy measurements at these wavelengths is the
confusion resulting from the thermal dust emission within our own
Galaxy. We consider the extent to which the Galactic dust emission can
be removed using multiwavelength informations from Herschel-SPIRE and
using Herschel complemented by Planck data on the same survey
field. Finally, we also consider how to optimize the area of a SPIRE
wide-field survey assuming a fixed observation time and consider the
extent to which information related to occupation number of FIRB
sources can be extracted.

The paper is organized as follows.  In the next Section, we discuss
the angular power spectrum of far-IR anisotropies based on a halo
model for the far-IR sources normalized to be roughly consistent with
luminosity functions of Lagache et al. (2003). We discuss issues
related to foreground confusion and the multi-frequency component
separation of Galactic dust in Section~3. In Section~4 we discuss the
applications of anisotropy measurements and outline the importance of
Herschel measurements at higher angular resolution than Planck.  The
latter only provides clustering measurements in the 2-halo part of the
anisotropy spectrum, while to extract some information on the source
distribution, clustering measurements at small angular scales
corresponding to the 1-halo term and observable with Herschel-SPIRE
are required.  When illustrating our calculations, we take
cosmological parameters from the currently favored flat-$\Lambda$CDM
cosmology with $\Omega_m=0.3$ and $\sigma_8=0.84$.

\section{Angular Power Spectrum}
\label{sec:fisher}

As mentioned in the introduction, to describe the FIRB anisotropy
power spectrum, we make use of an approach based on the halo
model. Using the Limber approximation (Limber 1954), the angular
power spectrum can be written as (Knox et al. 2001)
\begin{equation}
C_l^{\lambda \lambda'} = \int dz \frac{dr}{dz} \frac{a^2(z)}{d_A^2}
\bar{i}_\lambda(z) \bar{i}_{\lambda'}(z)
P_{ss}\left(k=\frac{l}{d_A},z\right) \, ,
\label{eq:cl}
\end{equation}
where $\rad$ is the conformal distance or lookback time from the
observer, $\da$ is the comoving angular diameter distance, and
$\bar{j}_\lambda(z)$ is the mean emissivity per comoving unit volume
at wavelength $\lambda$ as a function of redshift $z$ for sources
below a certain flux limit.  

Instead of intensity units, hereafter, we will work primarily in terms
of antenna temperature units ($\mu$K$_{RJ}$) with the conversion
factor for the angular power spectrum given as $\left({\partial
I(\lambda) \over\partial T}{\partial I(\lambda')
\over \partial T}\right)^{-1} = {(\lambda\lambda')^2 \over 4k_B^2}$.
We obtain $\bar{j}_\lambda(z)$ using the luminosity function models of
Lagache et al. (2003).  Note that the contribution to the IRB
intensity, at a given wavelength, is $I_\lambda = \int_0^{\infty} dz\;
\frac{dr}{dz} a(z) \bar{i}_\lambda(z)$ and can also be written as 
$I_\lambda= \int_0^{\infty} S (dN/dS) dS$ once luminosities are
converted to fluxes and $dN/dS$ is the differential number counts
obtained through a volume integral of the luminosity functions.

In Eq.~(\ref{eq:cl}), fluctuations in the source density field are
characterized by the three dimensional power spectrum $P_{ss}(k)=
P^{1h}(k) + P^{2h}(k)$. The two terms under the halo model are
clustering of FIRB sources in two different halos (2h) and clustering
within the same halo (1h), and given by (Cooray \& Sheth 2002):
\begin{eqnarray}
P^{2h}(k) &=&  \left[\int dM\; n(M)\; b(M) \frac{\langle N_{\rm t}(M)\rangle}{{\bar n}_g} u(k|M) \right]^2 P^{\rm lin}(k)\nonumber \\
P^{1h}(k) &=&  \int dM\; n(M)\; \frac{2 \langle N_{\rm s} \rangle \langle N_{\rm c} \rangle u(k|M)  + \langle N_{\rm s} \rangle^2 u^2(k|M)}{\bar{n}_g^2} \, ,
\label{eqn:pk}
\end{eqnarray}
respectively with the halo occupation number $\langle N_{\rm
t}(M)\rangle= \langle N_{\rm s} \rangle + \langle N_{\rm c} \rangle$.
Here, $u(k|M)$ is the normalized density profile in Fourier space,
$n(M)$ is the halo mass function, $b(M)$ is the halo bias relative to
the linear density field, and $\bar{n}_g$ is the number density of
sources.  As written, the 2-halo term with $P^{2h}(k)$ traces the
linear power spectrum scaled by a bias factor for these sources.

\subsection{Conditional Luminosity Functions}

In this paper, instead of the simple halo occupation number as
written above, we extend the halo model with conditional luminosity
functions to capture the luminosity distribution of galaxies in a
given dark matter halo (the CLF model; Cooray \& Milosavljevi\'c
2005; Cooray 2005; Yang et al. 2003; Yang et al. 2005).  The simple
halo occupation number treats all galaxies the same without allowing
for variations in the number counts with source luminosity or flux
while CLFs describe the number of galaxies in a given dark matter halo
mass as a function of the luminosity $\Phi(L|M)\equiv d\langle
N\rangle/dL$. Another motivation to consider a CLF model for FIRB
sources is the availability of preliminary models of the source
counts which assume the behavior of the redshift evolution of
far-IR source SED (e.g., Lagache et al. 2003). We make use of
the luminosity functions at 350 $\mu$m that were generated by Lagache
et al. (2003) as a function of the redshift, $\Phi^{\rm L03}(L,z)$,
and model CLFs to return a function comparable to these luminosity
functions once galaxy luminosities are associated with dark matter
halos.

The matching is done such that we require
\begin{equation}
\Phi(L,z) = \int dM dn/dM(z) \Phi(L|M,z) \, ,
\end{equation}
to be roughly consistent with  the redshift-dependent luminosity
functions of Lagache et al. (2003) by varying parameters related to $\Phi(L|M,z)$. We do not consider detailed models
through a likelihood analysis since Lagache et al. (2003) luminosity
functions are purely a phenomenological model of the source
distribution based on an assumed SED for galaxies with a
normalization to reproduce 60 $\mu$m IRAS luminosity function at low
redshifts. The model involves two types of far-IR sources described as
``normal'' galaxies and ``starburst'' galaxies. The normal galaxies
have a luminosity function of the Schechter form with a pure $(1+z)$
number density evolution out to a $z=0.4$ and a constant comoving
density beyond that to $z=5$. The luminosity function of the
starburst population and its evolution are more complicated
since they both involve the evolution in the density and in the
cut-off luminosity.  In Lagache et al. (2003), the redshift
dependences are taken to  be consistent with existing
observations so far at 15 $\mu$m and 850 $\mu$m. We do not reproduce
those details here but refer the reader to the paper by Lagache et
al. (2003).

To capture a consistent shape and redshift evolution from our CLFs, we
divide the source sample into normal and starburst galaxies and
consider models of the two populations separately. Following
well-known results in studies related to the galaxy distribution that
show differences in the properties of  central and satellite
galaxies in dark matter halos (e.g., Kravtsov et al. 2004; Berlind et
al. 2003), we subdivide the CLF into central and satellite
galaxies for both populations. These conditional functions are
written in the same manner that has been used to study galaxy
statistics at optical and near-IR wavelengths, with
\begin{eqnarray}
\Phi(L|M,z)&=&\Phi^{\rm cen}(L|M,z)+\Phi^{\rm sat}(L|M,z) \nonumber \\
\Phi^{\rm cen}(L|M,z)  &=& \frac{f_{\rm cen}(M,z)}{\sqrt{2 \pi} \ln(10)\sigma_{\rm cen} L} \times \nonumber \\
&& \quad \quad \exp \left\{-\frac{\log_{10} [L /L_{\rm c}(M,z)]^2}{2 \sigma_{\rm cen}}\right\}  \nonumber \\
\Phi^{\rm sat}(L|M,z) &=& A(M,z) L^{\gamma} \, .
\label{eqn:clf}
\end{eqnarray}
Here, $f_{\rm cen}(M,z)$ is a selection function introduced to account
for the efficiency of galaxy formation as a function of the halo mass,
given that the galaxy formation of low mass halos may be inefficient
and that not all dark matter halos may host a galaxy:
\begin{equation}
f_{\rm cen}(M,z) = \frac{1}{2}\left[1+{\rm erf}\left(\frac{\log(M)-\log(M_{\rm cen-cut}(z))}{\sigma}\right)\right] \\
, .
\label{eqn:fcm}
\end{equation}
In our fiducial description, we will take numerical values of  $M_{\rm cen-cut}= 10^{10}$ M$_{\rm sun}$ and $\sigma=0.5$ (Cooray 2005).

In  Equation~4, $L_c(M,z)$ is the relation between the
central galaxy luminosity of a given dark matter halo and its
halo mass, taken to be a function of the redshift, while $\sigma_{\rm
cen}$, with an assumed value of 0.25 to reproduce the shape of Lagache
et al. luminosity functions when $L > 10^{11}$, is the log-normal
dispersion in this relation.  For an analytical description of the
$L_{\rm c}(M,z)$ relation, we make use of the form suggested by Vale
\& Ostriker (2004) where this relation was established as 
appropriate for $b_J$-band galaxies by inverting the 2dFGRS luminosity
function.  We follow the same procedure by inverting the 60 $\mu$m LF
at low redshifts from IRAS data as well as 350 $\mu$m LFs as derived
by Lagache et al. (2003).  The relation is described with a general
fitting formula given by
\begin{equation}
\label{eqn:lcm}
L_{\rm c}(M,z) = L_0(1+z)^{\alpha} \frac{(M/M_1)^{a}}{[b+(M/M_1)^{cd(1+z)^{\eta}}]^{1/d}}\, .
\end{equation}
At 350 $\mu$m, for starburst galaxies, the parameters have values of
$L_0=2\times10^{12} L_{\sun}$, $M_1=6\times10^{12} M_{\sun}$, $a=4.0$,
$b=0.4$, $c=3.8$, and $d=0.23$. For other wavelengths this relation
can be simply scaled based on a spectral energy distribution such as
the one employed in Lagache et al. (2003).  To capture the redshift
dependence of the luminosity function, we set non-zero values for
$\alpha$ and $\eta$ with 0.05 and 0.1, respectively. We employ the
same analytical model for normal galaxies that appear at halo centers
with parameters $L_0=6\times10^{10} L_{\sun}$, $M_1=5\times10^{10}
M_{\sun}$, $a=4.0$, $b=0.6$, $c=3.9$, and $d=0.2$ and ignore the mild
redshift evolution at low redshifts in the Lagache et al. (2003)
model. This is not a concern for us since the anisotropies that will
be studied with Herschel-SPIRE will be dominated by the starburst
galaxy population while proper statistics of the normal galaxy
population will only be needed to understand the source counts at low
redshifts based on the resolved number counts.

In Figure~1, we plot a comparison of the $L_{\rm c}(M,z)$ relation at
$z=1$ for both normal and starburst galaxies. As shown and extracted,
based on this simple model, normal galaxies can appear as central
galaxies in halo down to smaller mass than the starburst galaxies. In
return, starburst galaxies that appear at halo center are brighter at
350 $\mu$m than normal galaxies for the same halo mass if
the halo mass is greater than 2.10$^{12} M_{\sun}$.

For satellites, the normalization $A(M)$ of the satellite CLF can be
obtained by defining $L_{\rm s}(M,z)\equiv L_{\rm tot}(M,z)-L_{\rm
c}(M,z)$ and requiring that $L_{\rm s}(M,z)=\int_{L_{\rm min}}^{L_{\rm
max}} \Phi^{\rm sat}(L|M,z)LdL$, where the
minimum luminosity of a satellite is $L_{\rm min}$.  In the luminosity
ranges of interest, with $\gamma$ between -0.5 to -1, CLFs are mostly independent of the exact value
assumed for $L_{\rm min}$ as long as it lies in the range
$(10^6-10^8)L_{\sun}$.  To describe the total far-IR luminosity of a
given dark matter halo we make use of the following phenomenological
form:
\begin{eqnarray}
L_{\rm tot}(M,z) = \left\{\begin{array}{ll}
L_{\rm c}(M,z)  & M \leq M_{\rm sat}\\
L_{\rm c}(M,z)\left(\frac{M}{M_{\rm sat}}\right)^{4} & M>M_{\rm sat}
\end{array}\right.
\label{eqn:ltot}
\end{eqnarray}
Here, $M_{\rm sat}$ denotes the mass-scale at which satellites begin
to appear in dark matter halos (taken to $\sim 10^{11}$ for normal
galaxies and $5 \times 10^{12}$ for starburst galaxies) with
luminosities as corresponding to those in the given sample of
galaxies. The power-law slope is fixed at 4, independently of the
redshift and consistent with total galaxy luminosity-cluster mass
relations at near-IR wavelengths (e.g., Lin \& Mohr 2004). Whether
such a relation holds for far-IR luminosity content of dark matter
halos may be testable with Herschel data in the same manner it has
been tested at optical and near-IR wavelengths using cluster catalogs.

Since we have divided the far-IR source population  into normal and
starburst galaxies, we have an additional freedom on how to distribute
these galaxies in dark matter halos and this freedom leads to large
degeneracies that cannot be simply separated from the luminosity
function alone, even if the luminosity function is available for both
source types separately. The clustering measurements, including
unresolved anisotropies, will make possible to correctly match the
mass scales and the relative distribution of normal and starburst
populations in dark matter halos in the same manner luminosity
functions and clustering studies have been used to identify relative
distributions of early- and late-type galaxies (such as the fraction
of central galaxies, that appear as late or early type, as a
function of the halo mass) at optical wavelengths. If we assume that a
fraction of $f_c(M,z)$ galaxies at halo centers are normal galaxies
and a fraction $f_s(M,z)$ of satellite galaxies are normal galaxies,
then the luminosity function of normal (n) galaxies can be written as
\begin{eqnarray}
&&\Phi^{\rm n}(L,z) = \int dM\, \frac{dn}{dM}(z)\, \nonumber \\
&\times& \left[f_c(M,z) \Phi_{\rm c}^{\rm n}(L|M,z) + f_s(M,z) \Phi_{\rm s}^{\rm n}(L|M,z)\right] \, .
\end{eqnarray}

The luminosity function of starburst galaxies follows by taking
$1-f_c(M,z)$ and $1-f_s(M,z)$ fractions and integrating with CLFs as
defined for starburst galaxy population. Given that we have large
degeneracies in how to distribute the population, we take the simple
approach that in a given halo, we can take the fraction to be same for
satellites and centrals (in practice since only one central galaxy is
present, this would mean that some halo centers have central galaxies
while others have starbursts) and let $f_c(M,z)=0.1$, independent of
redshift and above the mass at which starburst galaxies appear in halo
centers, while this fraction is 1 below the mass scale at which
starbursts appear (this mass scale is determined based on a simple
match to starburst luminosity functions of Lagache et al. 2003). The
exact halo mass and redshift dependences would be some of the
interesting parameters that can be potentially extracted from data if
adequate statistics are available.  The same parameters can be
investigated from semi-analytical models of galaxy formation with a
focus on the far-IR luminosities and we encourage such studies with
existing models.

In Figure~2, we illustrate the comparison between our model, which
uses the dark matter halo mass as the starting point to model the
galaxy distribution, and Lagache et al. (2003) LFs at 350 $\mu$m at
$z=1$, which uses a SED and the 60 $\mu$m LF at low redshifts as
a way to build an evolutionary model of source counts with parameters
matched to existing data.  In Figure~3, we show the normalized redshift
distribution of sources (both normal and starburst galaxy types),
$\int dz\, n(z) =1$, with fluxes below 100 mJy. As shown here, the redshift
distribution is such that the numbers peak at $z > 1$, but extends to higher
redshifts with a very slow decrease  when $z> 3$.

While we have not performed a detailed model fit, which we think is
premature given the limited data at these far-IR wavelengths, we have
varied parameters to get a reasonable consistency between two
prescriptions in the literature to model IR source counts.  More
importantly, our description is outlined in a way that, we think, will
allow to extract a model based on future measurements of both
luminosity functions and clustering at far-IR wavelengths.

\subsection{Far-IR Source Occupation Numbers}

By integrating CLFs over luminosity down to a fixed luminosity, we can
recover the halo occupation number through
\begin{equation}
\langle N(M) \rangle = \int dL \Phi(L|M) \, .
\end{equation}
In Figure~4, we illustrate the occupation number at $z=1$ for both
normal and starburst populations (and divided to central and satellite
galaxies in both cases) with $L_{\rm max}$ corresponding to sources
with a maximum flux below 100 mJy at 350 $\mu$m. The combined
occupation number cannot be simply described by a power-law though the
parameter $\gamma$ describing the satellite CLF luminosity
dependence is related to the power-law slope $\beta_s$ when occupation
number of satellites is described such that $\langle N_s(M) \rangle
\propto M^\beta_s$ at the high mass end.  Note that the central galaxy
occupation number is always one above some mass scale.  For sources
fainter than 100 mJy at $z=1$ at 350 $\mu$m, the minimum mass at which
normal galaxies appear is about few times $10^{10}$ M$_{\sun}$, while
the mass scale at which starburst appear is few times 10$^{11}$
M$_{\sun}$. When calculating the power spectrum, we integrate the CLFs
over the assumed cut-off luminosity to extract the necessary
occupation numbers for both central and satellite galaxies.

While we have provided a bit complicated model, so as to illustrate
how the exact modeling can be done once luminosity functions are
established at far-IR wavelengths at several different redshifts
beyond the low-redshift short wavelength LFs now available from
surveys with IRAS, for clustering studies, the behavior of the
occupation number as suggested by CLFs can be captured with a model of
the form $\langle N_{\rm t}(M,z) \rangle = 1 + \langle N_{\rm
s}(M)\rangle$ when $M > M_{\rm min}(z)$ and 0 otherwise, with the
assumption of a central FIRB source in each halo ($\langle N_{\rm
c}(M)\rangle=1$) above some mass scale and a power-law distribution of
satellites with $\langle N_{\rm s}(M)\rangle = (M/M_{\rm
min})^\beta$. The value of $M_{\rm min}$, for example, can be readily
extracted from occupation numbers shown in Figure~4, or if one
needs to vary to a different flux, based on the matching between
luminosity and mass from the $L_c(M)$ relation shown in Figure~1, for
example.  Instead of parameters in the CLF, we will show the extent to
which we can extract a parameter such as $\beta$ from the 1-halo term
of the anisotropy power spectrum given that the small scale clustering
is strongly sensitive to the statistics related to satellite galaxies.

\begin{figure}[!t]
\begin{center}
\includegraphics[width=8.5cm]{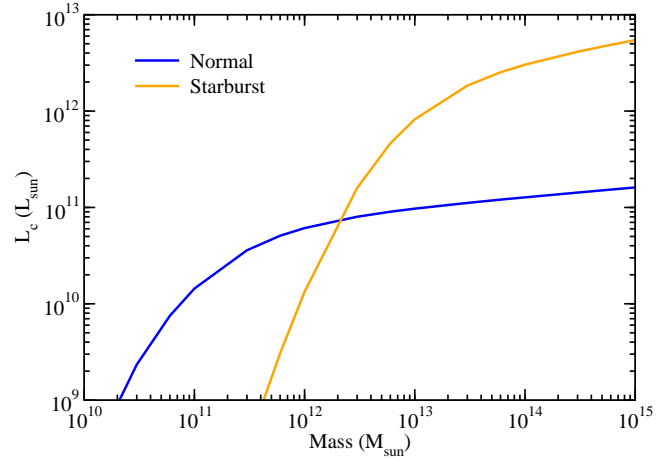}
\caption{The relation between the 350 $\mu$m luminosity of a central 
galaxy in a dark matter halo as a function of the halo mass (at a
redshift of one). We also show the expected relation between the 350
$\mu$m luminosity of a ``normal'' galaxy at a redshift of 0.3 as a
function of the halo mass. In our modeling, we are assuming that 90\%
of the central galaxies in halos roughly above 10$^{12}$ M$_{\sun}$
are of the starburst type while at the low mass-end of dark matter
halos, all central galaxies are normal galaxy types, in the
prescription of Lagache et al. (2003).}
\label{fig:lcm}
\end{center}
\end{figure}

\begin{figure}[!t]
\begin{center}
\includegraphics[width=8.5cm]{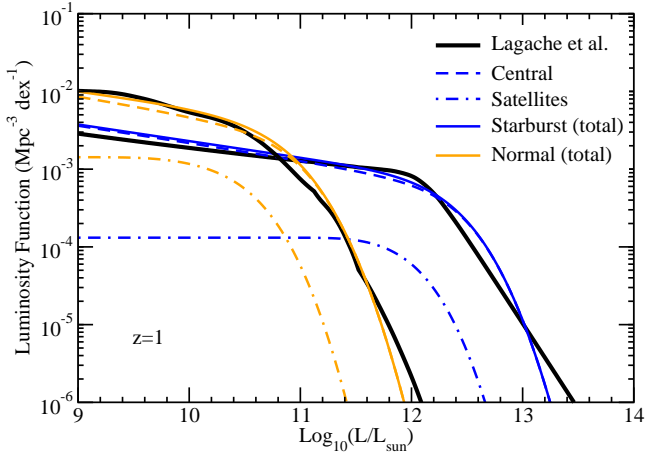}
\caption{The luminosity function of far-IR sources at $z=1$. 
In black lines, we show the LFs from the source evolution model of
Lagache et al. (2003) subdivided to normal and starburst galaxy
populations (thick solid lines). The thin lines are the LFs predicted
from the CLF halo model described here, with long dashed lines showing
the contribution from central galaxies and the dot-dashed lines
showing the contribution from satellites in each of the two types. We
obtain a rough agreement between the two different descriptions.  With
detailed measurements of the LFs with various surveys using Herschel,
it is likely that an approach such as the one presented here can be
used to establish the underlying source model.}
\label{fig:lf}
\end{center}
\end{figure}

\begin{figure}[!t]
\begin{center}
\includegraphics[width=6.5cm,angle=-90]{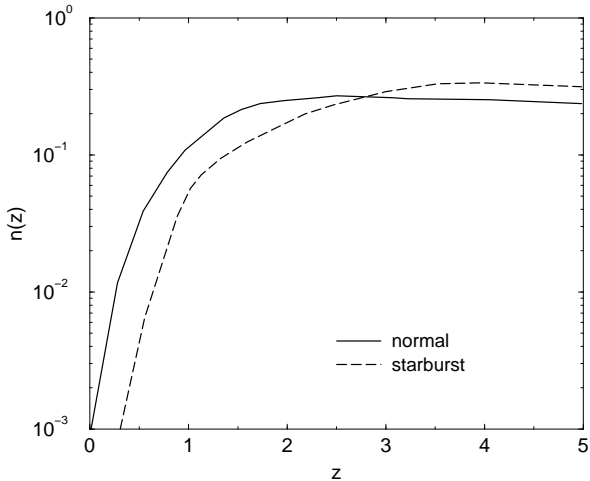}
\caption{The predicted normalized redshift distribution for normal 
and starburst galaxies at 350 $\mu$m, with fluxes less than 100 mJy
over all redshifts, based on the CLF halo model presented here.}
\label{fig:nz}
\end{center}
\end{figure}

\begin{figure}[!t]
\begin{center}
\includegraphics[width=8.5cm]{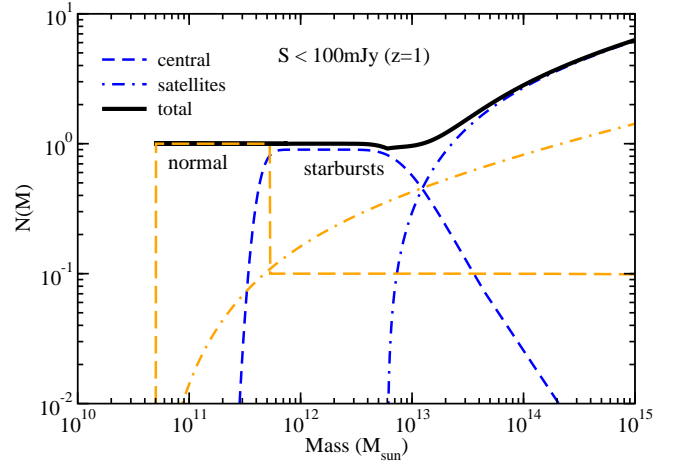}
\caption{The far-IR halo occupation number by integrating CLFs with a maximum luminosity corresponding to a flux of 
100 mJy and at for sources at $z=1$. The occupation number is divided
to normal and starburst populations and to central (long-dashed lines)
and satellite galaxies (dot-dashed lines) in each of the far-IR source
types under the classification of Lagache et al. (2003). The thick
solid line is the total halo occupation number.  The central galaxies
of normal type are mostly in low mass halos while central galaxies of
starburst type are in halos at the high mass end. The decrease in the
central occupation number of starburst galaxies below 1 when $M >
10^{13}$ M$_{\sun}$ is due to the fact that we have only shown the
occupation number for sources with fluxes below 100 mJy here.}
\label{fig:nm}
\end{center}
\end{figure}

\begin{figure*}[!t]
\begin{center}
\includegraphics[width=10cm]{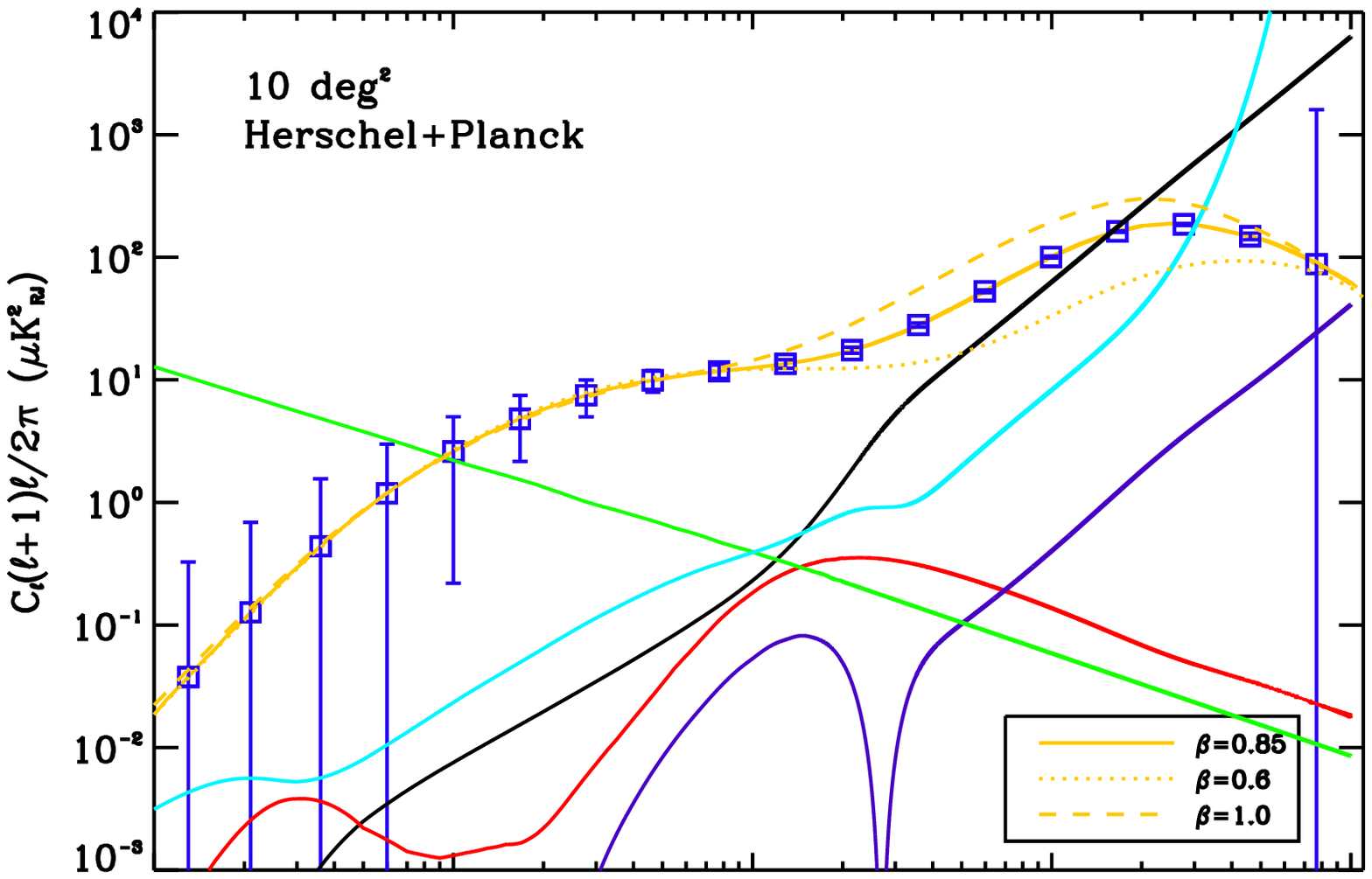} 
\hspace{-2.0cm}
\includegraphics[width=10cm]{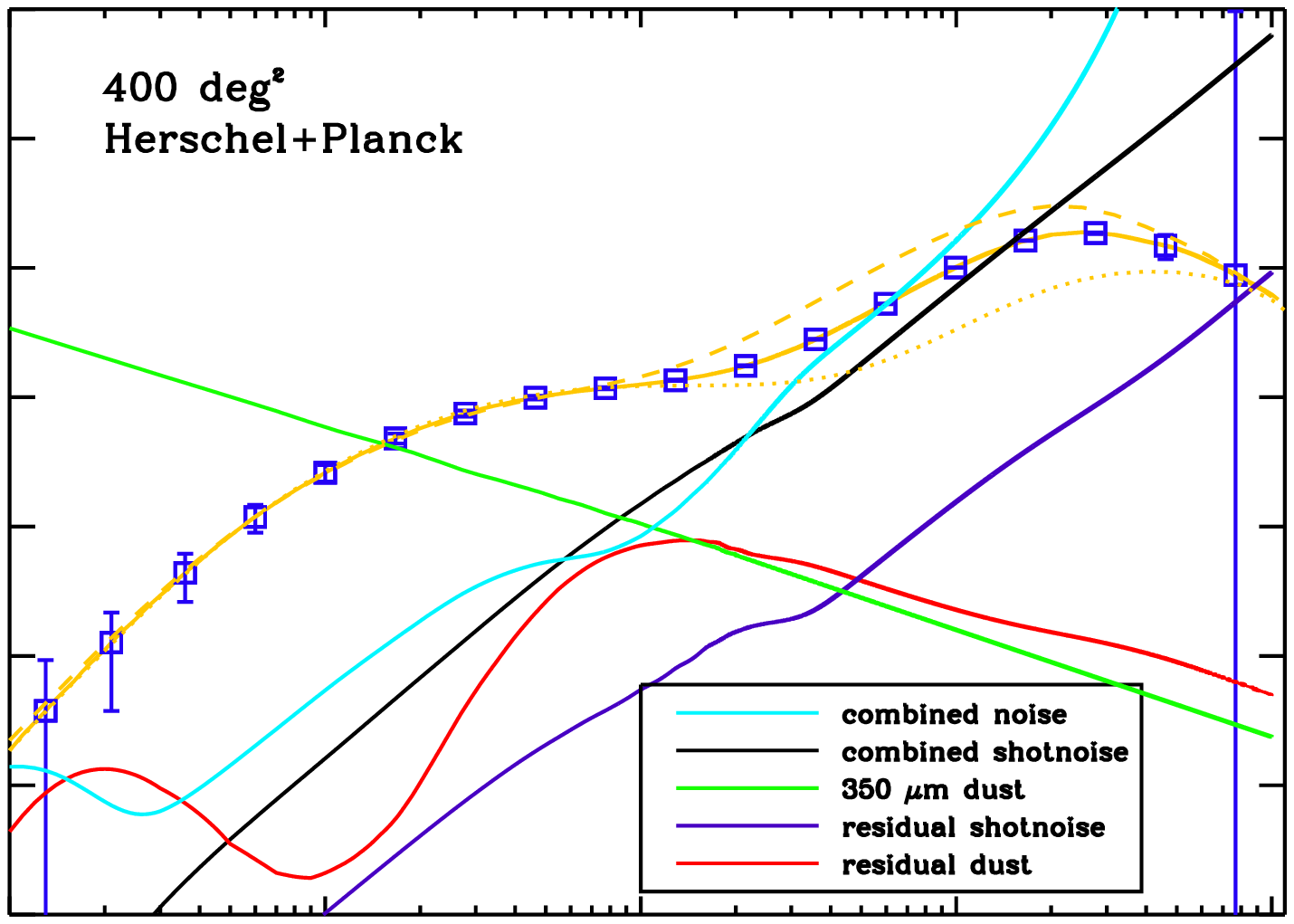}\\
\vspace{-1.3cm}
\includegraphics[width=10cm]{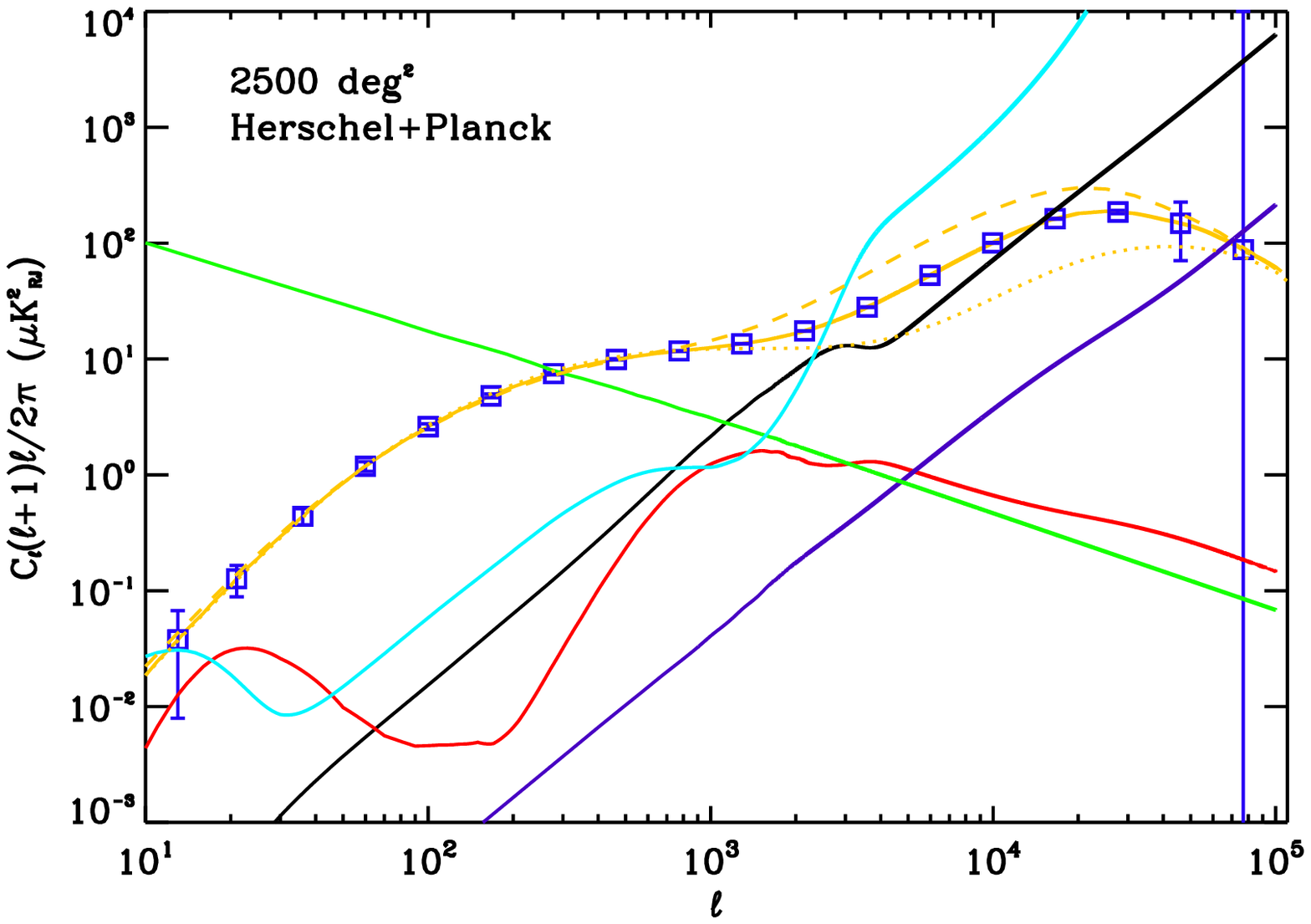}
\hspace{-2.0cm}
\includegraphics[width=10cm]{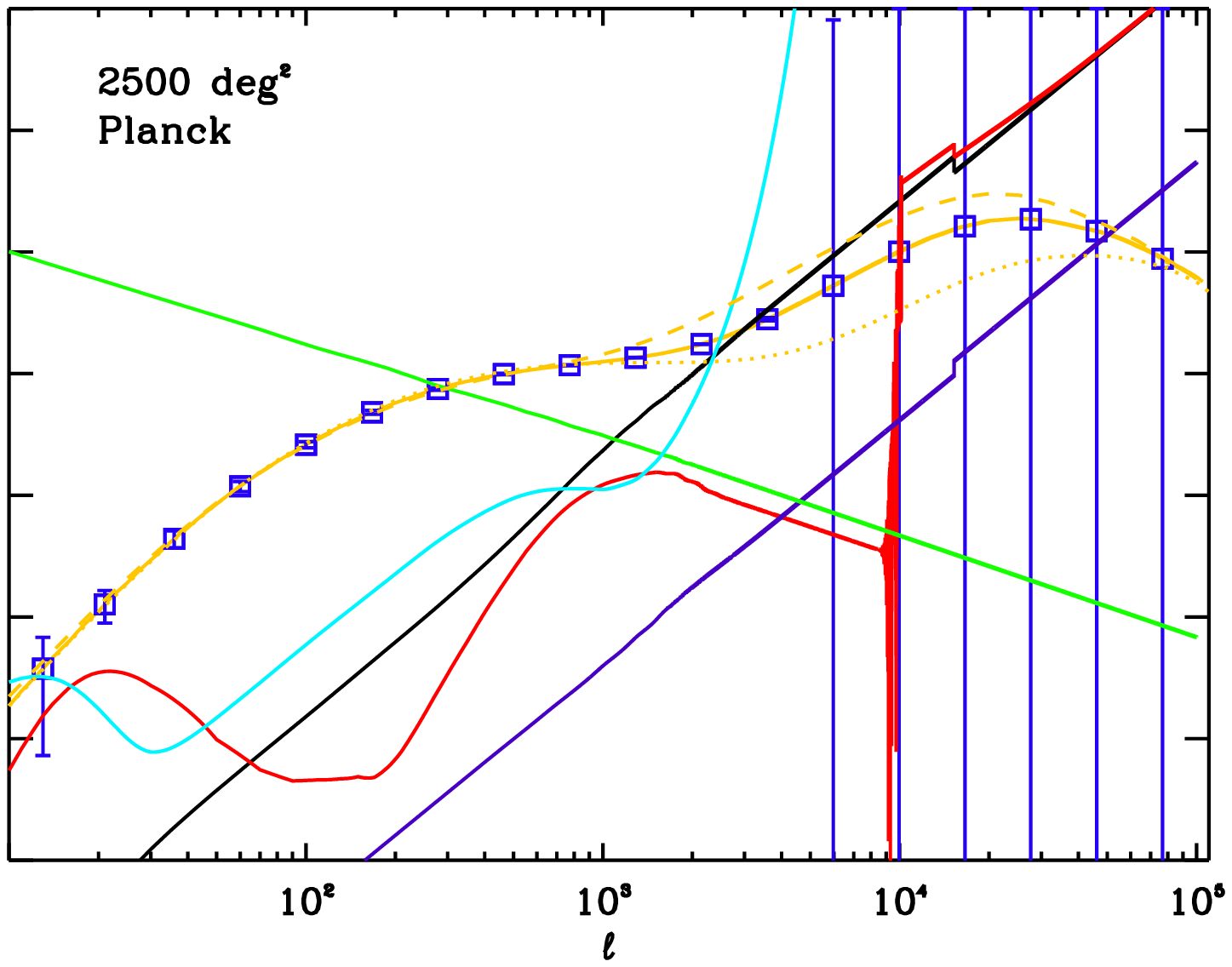}
\caption{Dust removal results for a 10, 400 and 2500 deg$^2$ area
with Planck and Herschel and for a 2500 deg$^2$ area with Planck
around the ELAIS-S1 field (from top to bottom and left to right). All
spectra are in antenna temperature, $\mu$K$_{RJ}^2$, units at 350
$\mu$m. The foreground and shot-noise residual are respectively the
solid red and purple lines. The green solid line is the Galactic dust
power spectrum at 350 $\mu$m. The solid light blue and black lines are
respectively the combined detector noise (once the optimal frequency
mixing is applied) and the combined shot-noise. The FIRB clustering
contribution is shown as orange solid, dotted, dashed lines for
$\beta$ of 0.85, 0.6, 1 respectively. The blue squares represent the
binned theoretical FIRB power spectrum with its error
(${\Delta\ell\over\ell}=0.5$).}
\label{fig:residsep}
\end{center}
\end{figure*}

\subsection{Far-IR Source Clustering Bias}

In general source clustering at large angular scales can be described with the linear matter power
spectrum scaled by a constant, scale-free bias factor such that
\begin{equation}
P_{ss}(k) \approx b^2_s P^{\rm lin}(k) \, .
\end{equation}
In terms of the CLF halo model, this large scale source bias can be written as a combination of the bias of normal and
starburst galaxies by noting that the large-scale bias factor of, for example, normal galaxies as a function of
source luminosity and the redshift is
\begin{eqnarray}
&&b^{\rm n}(L,z) = \int dM\, \frac{dn}{dM}(z)\, b_{\rm halo}(M,z)\nonumber \\
&\times& \frac{\left[f_c(M,z) \Phi_{\rm c}^{\rm n}(L|M,z) + f_s(M,z) \Phi_{\rm s}^{\rm n}(L|M,z)\right]}{\Phi^n(L,z)} \, .
\end{eqnarray}
The total galaxy bias, as necessary for unresolved anisotropy clustering
measurements, can be calculated by replacing $\Phi(L|M,z)$ with the sum of
both normal and starburst CLFs and replacing the LF for normal galaxies with the total. 

When separately considered, given that normal galaxies are found mostly in halo centers at the low mass end, the predicted bias factors are generally at the level of $\sim 1$ or slight below one. The starburst galaxies, however, are mostly at halo centers at the high mass end and their bias factors are expected to be larger than 1. Thus, the starburst population is expected to be
both strongly clustered and to have both large correlation lengths or bias 
factors. Using the CLF halo model, if the  clustering spectrum 
of resolved sources at $z \sim 1$ with fluxes
above 100 mJy is measured, we find that the bias factors will be expected 
to be order 2.0. Such a large bias factor, or equivalently a large correlation length,
is consistent with some of the limited suggestions in the literature that far-IR sources with
bright fluxes are strongly clustered (e.g., Blain et al. 2004).
At the same flux cut, the unresolved anisotropies are
expected to have a bias factor of the order 1.1. This value is substantially
below the bias factor of resolved sources at the same flux cut since
unresolved anisotropies are dominated by sources that are substantially fainter than the sources just below the flux cut.

\subsection{Shot-noise power spectrum}

In addition to the clustering signal, at small angular scales, the
finite density of sources leads to a shot-noise type power spectrum in
the IRB spatial fluctuations. This shot-noise can be estimated through
number counts such that $C_l^{\rm shot} = \int_{0}^{S_{\rm cut}} dS\,
S^2 (dN/dS)$ when $S_{\rm cut}$ is the flux cut off value related to
the removal of resolved sources.We use the confusion noise as
estimated by Hspot\footnote{http://herschel.esac.esa.int/ao\_kp\_documentation.shtml}
for Herschel, 25, 29, 24.5 mJy (5$\sigma$) at 250, 350 and 500 $\mu$m
and by Negrello et al. (2004) for Planck, 29.2, 115, 323 and 705 mJy
(5$\sigma$) at 217, 353, 545 and 857 GHz.

\section{Foreground Separation}

We model the Galactic dust with the model 8 (two temperature model) of
\cite{Finetal99} and maps from \cite{SFD98} (cleaned IRAS with a
calibration obtained on DIRBE with an effective angular resolution of
6 arcminutes) over the frequency range of 217 GHz to 1200 GHz (250
$\mu$m to 1.2 mm). We fit with a power law the power spectrum of the
Galactic dust of a 4900 deg$^2$ area around ELAIS-S1 field at each
frequency. We use the variance of the signal in our smaller fields to
scale down the fitted power spectrum to these smaller fields. The FIRB
anisotropy power spectrum modeled at 350 $\mu$m following the halo
model above is interpolated to other wavelengths using the mean
spectrum from COBE/FIRAS (Fixsen et al. 1998) as
\begin{equation}
I_\nu=\tau_0(\nu/\nu_0)^\alpha B_v(T) \ ,
\end{equation}
with $\tau_0=(1.3\pm 0.4)\times 10^{-5}$, $T=18.5 \pm 1.2$K and $\alpha=0.64 \pm 0.12$.

To study the extent to which foreground dust confusion can be reduced
we used the cleaning technique outlined in \cite{Tegetal03}, where
multifrequency maps from WMAP first-year data were used to produce the
so-called TOH foreground-cleaned CMB map.  The technique recommends
taking a linear combination of observed $a_{\ell m}$'s in each
frequency band $i$, $a_{\ell m}=\sum_{freq=i}w^i_{\ell}a^i_{\ell m}$,
with weights $w_i$ chosen to minimize foreground contamination from
Galactic dust.  We decompose the signal at each frequency as
$a^i_{\ell m} = u^i_{\ell m} + f^i_{\ell m} + n^i_{\ell m}$ where $u$,
$f$, and $n$ stand respectively for unresolved FIRB, foregrounds (CMB
and Galactic dust), and noise.  We then minimize the resulting power
spectrum of extragalactic FIRB fluctuations, assuming the frequency
spectrum of this component follows that of \cite{Fixsenetal98} through
$\langle|a_{\ell m}|^2\rangle= {\bf w_\ell}^T {\cal C} {\bf w_\ell}$
using weights under the constraint $w_\ell^T \cdot {\bf e}=1$, where
${\bf e}$ is a column vector with the relative amplitude of the FIRB
spectrum.

The above condition allows to sum the different frequencies without
reducing the FIRB component, and permits to optimally subtract the
Galactic dust (CMB itself as a foreground only makes a minor impact at
frequencies above 500 GHz).  Here, the ${\cal C}^{ij}_\ell$ matrix
represents $\langle (a^i_{\ell m})^\dagger a^j_{\ell m}
\rangle$.  As derived in \cite{Tegetal03}, the weights that minimize the power
$\langle|a_{\ell m}|^2\rangle$ are
\begin{equation}
{\bf w_\ell} = \frac{ {\cal C}^{-1} {\bf e}}{e^T {\cal C}^{-1} e} \, .
\end{equation}
The estimate of residual dust level with this method is clearly
optimistic since the FIRB spectrum is poorly known.  We will quantify
the impact of the uncertainty related to the spectrum in terms of an
overall uncertainty in the cumulative signal-to-noise ratio for
detection of FIRB fluctuations in the presence of Galactic dust. To
quantify our results, we center our simulation around ELAIS-S1 (RA-DEC
: 7.8,-44.2), which is a preferred region known to be least
contaminated with cirrus even after considering a large region around
this field.

In our calculations we assume an uncorrelated noise power spectrum, 
with $\langle n_{\ell m}^i n_{\ell' m'}^j \rangle = \delta_{ij}
\delta_{mm'} \delta_{\ell \ell'} N_\ell'$, between SPIRE bands. 
For $N_l'$, we take 1$\sigma$ noise levels for a SPIRE 10 deg$^2$
survey with a 1000 hour integration: 1.6, 2.1 and 1.7 mJy at 250, 350,
500 $\mu$m (obtained from HSpot) and then degrade as the survey area
is increased.  To combine Herschel data with Planck, we make use of
217, 353, 545 and 857 GHz channels of Planck HFI with equivalent noise
of 13.4, 25.2, 48.4 and 55.4 mJy
\footnote{http://www.rssd.esa.int/Planck}. Since wide-field
Herschel-SPIRE maps are raster-scanned, 1/f-noise impact measurements
of large angular scale fluctuations. We model this by modifying the
overall noise spectrum to be $N_\ell = N'_\ell(1+\ell_{\rm
knee}/\ell)$ (e.g., Crawford 2007) and take $\ell_{\rm knee}=10^3$
corresponding to a 1/f knee at a frequency of 100 mHz with a scan rate
of 60''/sec.  While we include 1/f-noise, to effectively remove it
requires two passes of the same field in orthogonal
directions. Generally, this requires that for a given area, one spend
twice as long than when a survey is conducted for point source
detections only. Given the maximum scan rate, for a fixed integration
time, there is also a maximum area one can cover, but we have ignored
this restriction here when estimating the signal-to-noise ratio as a
function of the sky area as the maximum scan rate is not yet
established.  We neglect Planck 1/f-noise residual, since we
believe it would not be as strong as Herschel one on our scales of
interest due to Planck's faster scanning speed.

\begin{figure}[!t]
\begin{center}
\hspace{-0.7cm}\includegraphics[width=9.5cm]{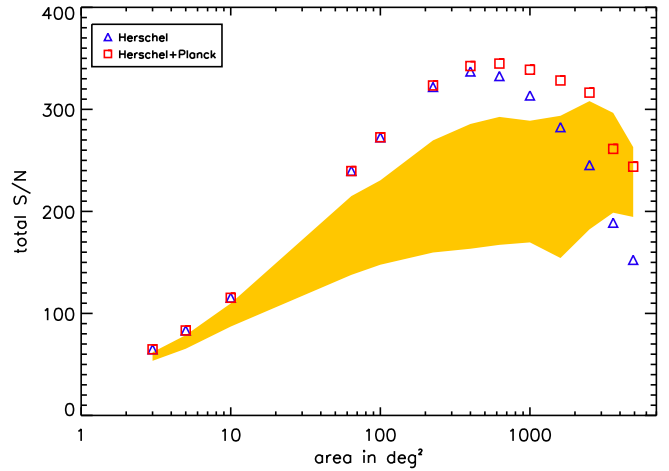}
\caption{Total signal-to-noise on the estimate of the FIRB clustering power
spectrum. We took into account the instrumental noise, the cosmic
variance, and the foreground and shot-noise residuals for
different sky coverage (from 3 to 5000 deg$^2$). The blue triangles
and red squares represent a ``clean'' area selected from SFD98 dust
map and centered around ELAIS-S1 field using Herschel and
Herschel+Planck, respectively. In each selected survey size, we assume
a total integration time of 1000 hours with Herschel-SPIRE, while the
integration time for Planck is independent of the survey area given
that Planck data is an all-sky survey be design.  The solid orange
area represents the signal-to-noise ratio achievable when FIRB
spectrum is taken to be uncertain at the $\pm$ 1 $\sigma$ level of
Fixsen et al. (1998) analytical model (see, Eq.~3).}
\label{fig:stonvsarea}
\end{center}
\end{figure}

\section{Results \& Discussion}

Figure ~\ref{fig:residsep} shows examples of the FIRB power spectrum
estimated on a 10, 400, and 2500 square degree fields around a very
low foreground Galactic dust field centered around ELAIS-S1
using the 3 SPIRE bands (corresponding to central frequencies of 577,
833 and 1200 GHz) with a 1000 hours integration time and 4 of
Planck frequency bands (217, 353, 545, 857 GHz, assuming the 14 month
survey).  According to figure ~\ref{fig:residsep}, these surveys can
measure accurately the power spectrum of FIRB on scales smaller 
than 30 arcminutes ($\ell \simeq 400$), however, the smallest scales
might be biased by the confusion noise coming from the shot-noise
term. Larger scales are not very well measured by the 10 deg$^2$
survey due to a large remaining cosmic variance. On the other hand,
the Galactic dust residuals are much smaller for the smaller area
survey (typically a factor 5 to 10 in power).

\subsection{Optimal survey area}

In addition to 3 field sizes highlighted in Fig.~\ref{fig:residsep},
it could be that the angular power spectrum might be measured at a
more significant level if the field size is optimized for these
measurements given a finite integration time.  We therefore compute
the noise level and the Galactic dust level in the FIRB power spectrum
estimate for different field size assuming a total of 1000 hour
integration time for Herschel/SPIRE observations.  To show
quantitatively how well different field sizes are measuring clustering
of the unresolved component, we computed the total signal to noise
(optimal sum on all the mode $\ell$), with the signal $C^{\rm
clus}_\ell$ being the FIRB power spectrum due to clustering, and the
noise being the sum of the instrumental noise ($N_\ell$) variance, the
cosmic variance and the residual Galactic dust and shot-noise :
\begin{equation}
S/N=\sqrt{\sum_{\ell=\ell_{min}}^{\ell_{max}}\left({C^{\rm clus}_\ell \over
X_\ell}\right)^2}\,  ,
\end{equation}
where
\begin{equation}
X_\ell=(N_\ell+C^{\rm FIRB}_\ell)\sqrt{2 \over (2\ell+1)f_{\rm sky}}+R_\ell \, ,
\label{eqn:xl}
\end{equation}
and $C^{\rm FIRB}_\ell=C^{\rm clus}_\ell+C_\ell^{\rm shot}$ and
$f_{\rm sky}$ is the fraction of sky covered. The residual shot-noise
is taken as the combination of the 1 $\sigma$ uncertainties computed
by a Fisher matrix analysis (Tegmark et al. 1997) at each frequency
for each instrument (Planck and Herschel/SPIRE), given a model where
dust and FIRB clustering power spectrum shape are known. These
residual shot-noise estimates are probably optimistic, especially for
Planck since its angular resolution does not allow to measure the
small angular scale ($\theta < 0.5'$) where the shot-noise term
dominates.

\begin{figure}[!t]
\begin{center}
\includegraphics[width=9.0cm]{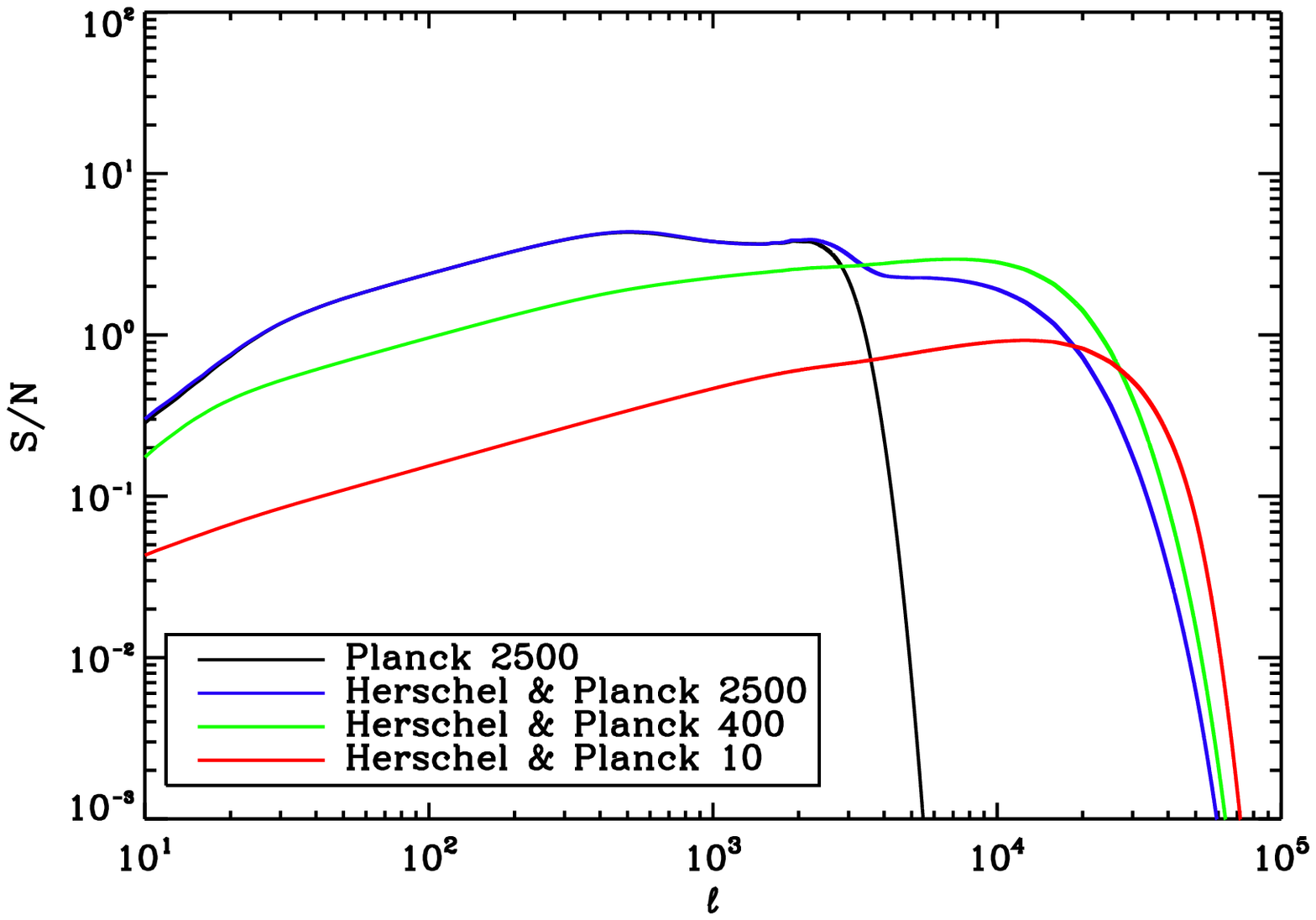}
\includegraphics[width=9.0cm]{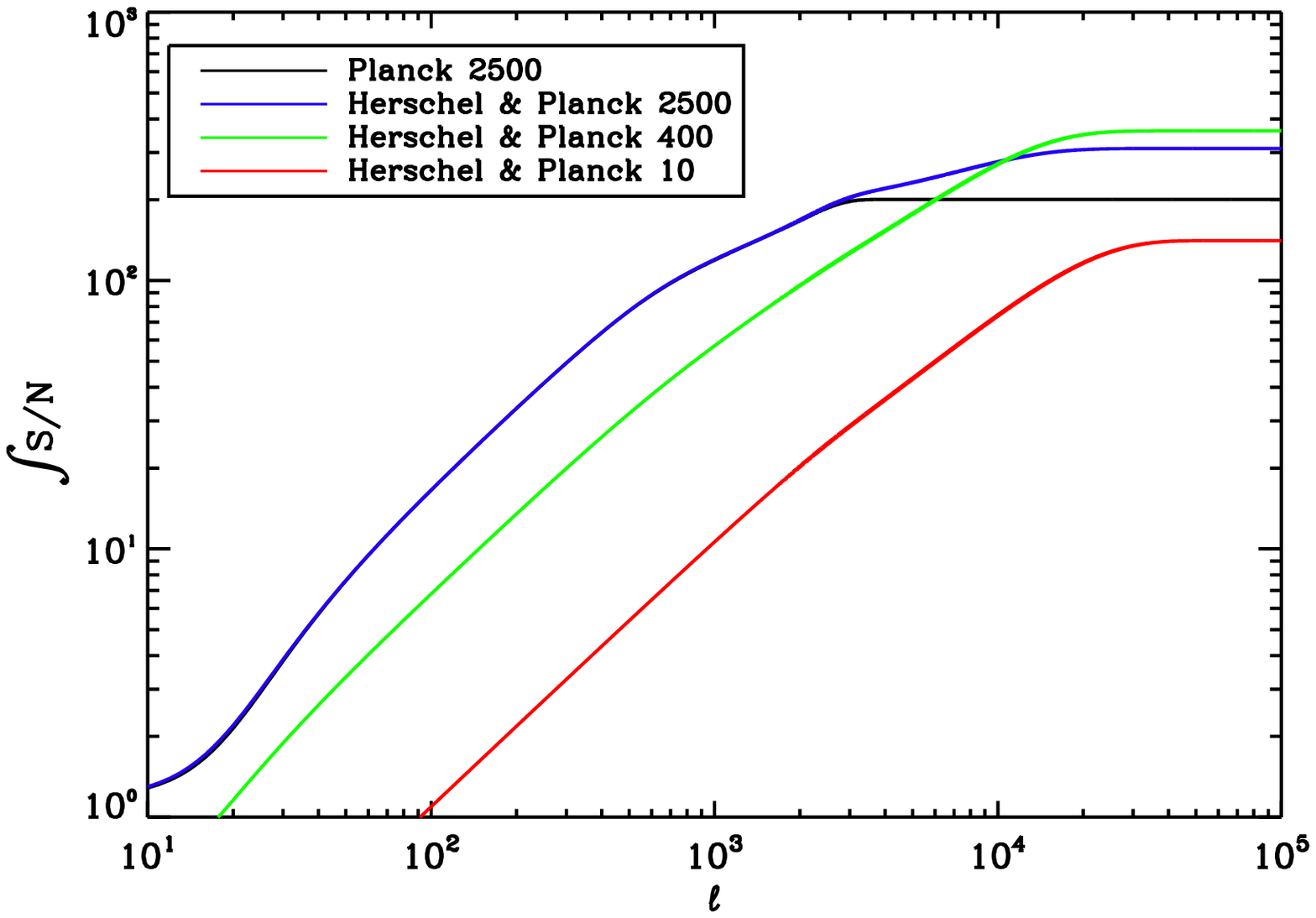}
\caption{The signal-to-noise ratio on the estimate of the FIRB clustering power
spectrum as a function of the multipole. {\it Top:} The
signal-to-noise as a function of the multipole. {\it Bottom:} The
cumulative signal-to-noise ratio as a function of the multipole. As
illustrated, Planck (HFI) can make higher signal-to-noise clustering
measurements of FIRB anisotropies, though the measurements are limited
to multipoles of less than about 4000. The combination of Herschel and
Planck over 400 deg$^2$, with Herschel imaging data in an integration
of 1000 hours, has the highest overall signal-to-noise ratio for the
clustering measurements.}
\label{fig:sn}
\end{center}
\end{figure}

The results obtained are summarized in
Figure~\ref{fig:stonvsarea}. The signal-to-noise ratio increases with
the area covered from 65 with 3 deg$^2$ to 360 with 400-600 deg$^2$
due to the decrease of the cosmic variance, and then decreases due to
the increase of the Galactic dust residual and the lower observation
depth. Adding the Planck channels at 217, 353, 545 and 857 GHz does
not change the overall sensitivity if the used Planck field is smaller
than 400 deg$^2$, but they increase substantially the sensitivity for
field of larger size. Figure~7 shows that Planck sensitivity dominates
at large scale ($\ell < 3000$) for field above 2500 deg$^2$, but that
a smaller Herschel field around 400 $deg^2$ is more optimal at
measuring smaller scale ($\ell > 3000$), whereas fields of few tens
square degree lose a lot of sensitivity in the clustering part
($1000<\ell<10000$) of the FIRB power spectrum due to the cosmic
variance. The ``optimal'' area (400 deg$^2$) corresponds to an
observation depth of about 51, 66, and 54 mJy (5 $\sigma$, only
instrumental) at 250, 350, 500 $\mu$m, we believe this ``optimal''
depth to be robust to a change in integration time, given our
shot-noise level estimates.

As discussed in Section~3, to estimate the confusion associated with
Galactic dust due to the uncertain extragalactic FIRB spectrum, we
vary the spectrum based on uncertainties in \cite{Fixsenetal98}
spectrum from COBE/FIRAS by drawing 100 simulations assuming
uncertainties in the parameters describing the spectrum are Gaussian
distributed (see, Eq.~3). Again, we separate the Galactic dust from
the FIRB and compute the total signal-to-noise ratio.
Figure~\ref{fig:stonvsarea} (orange area) shows that the sensitivity
to the clustering is degraded by 10\% to 35\% on average and
that the uncertainty on the FIRB spectrum generates a 10\% to 30\%
uncertainty on the total sensitivity. Even with these
uncertainties, the 400 deg$^2$ survey remains more sensitive than the
few ten square degree surveys. While uncertainties in foreground
emission largely impacts the overall signal-to-noise ratio for a
detection of FIRB anisotropy spectrum, an anisotropy study in a 400
deg.$^2$ field is still important given the limited knowledge we have
on the unresolved component that accounts for up to 90\% of the
background light at 350 $\mu$m.

\begin{figure}[!t]
\begin{center}
\hspace{-0.5cm}
\includegraphics[width=7.cm,angle=-90]{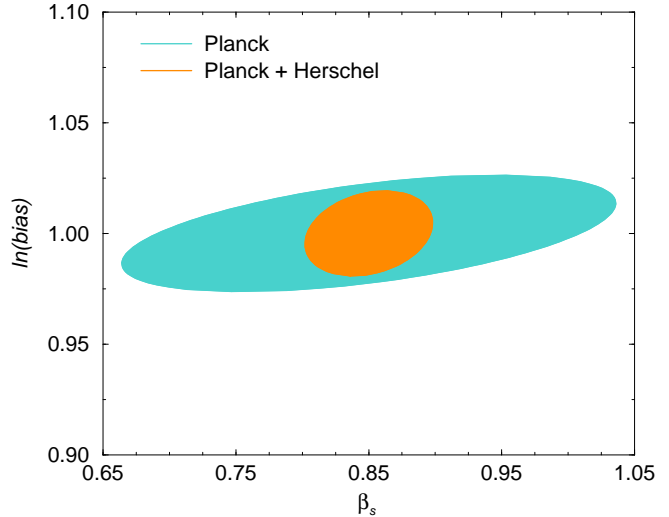}
\caption{Constraints on $\beta_s$ and the source bias factor from
unresolved FIRB clustering measurements with Herschel-SPIRE and
Planck. The ellipse describe the 95\% confidence level errors after
marginalizing over 2 additional parameters related to the uncertain
redshift distribution of sources.  As shown, Planck measurements (over
2500 deg$^2$) alone do not establish information captured in the
1-halo term, such as the satellite occupation number, while with
Herschel, small angular scale clustering is adequately measured to
properly establish statistics related to far-IR sources, such as the
slope of the occupation number. While we have only considered
clustering of unresolved anisotropies here, in a large-are survey, the
combination of LFs, resolved source clustering, and unresolved
anisotropies can be used to establish an improved model with
parameterization based on the CLFs.}
\label{fig:ellipse}
\end{center}
\end{figure}

\subsection{Astrophysical Information in Unresolved Anisotropies}

To study the extent to which these anisotropy measurements can be used
for astrophysical studies, we considered extraction of halo model
parameters.  For this, we assume a fiducial model for the source
distribution and allow variations in certain parameters related to
this model to study how the likelihood changes. This is done by
constructing the Fisher matrix,
\begin{equation}
{\bf F}_{ij} = -\left< \partial^2 \ln L \over \partial p_i \partial p_j \right>_{\bf X} \, ,
\end{equation}
where $L$ is the likelihood of observing a data set ${\bf X}$ given
the true parameters $p_1 \ldots p_n$. Since the variance of an
unbiased estimator of any parameter $p_i$ cannot be less than the
Cramer-Rao bound captured by $({\bf F}^{-1})_{ii}$, the Fisher matrix
quantifies the best statistical errors on parameters possible with a
given data set (Tegmark et al. 1997). We refer the reader to Knox et
al. (2001) for a prior application of the Fisher matrix to study how
well far-IR background anisotropies can be used to establish
properties of the source distribution.

For the clustering of unresolved anisotropies, the
Fisher matrix becomes
\begin{equation}
{\bf F}_{ij} = \sum_{\ell=\ell_{\rm min}}^{\ell_{\rm max}} \frac{1}{X_\ell^2}
{\partial C_\ell^{\rm clus} \over \partial p_i}
        {\partial C_\ell^{\rm clus} \over \partial p_j}\, , 
\label{eqn:Fisher}
\end{equation}
where $X_\ell$ follows from Equation~\ref{eqn:xl}. To establish the
relative importance of Herschel given Planck high frequency
observations and to get an order-of-magnitude estimate on how well
unresolved anisotropies can be used to extract some information on the
underlying source distribution, we consider a model with four
parameters involving large-scale bias factor, small scale occupation
number captured by the power-law slope $\beta_s$, and two parameters
to describe the redshift distribution of unresolved sources with
fluxes fainter than the point source detection.  Motivated by the
redshift distribution predicted by the CLF halo model and shown in
Fig.~\ref{fig:nz}, we parameterize the redshift distribution with a
quadratic function that is zero at $z=0$.  We calculate the Fisher
matrix by varying these four parameters and marginalize over the
uncertain redshift distribution by projecting the 4-dimensional Fisher
matrix to two dimensions involving bias and the power-law slope of the
occupation number.

In Figure~\ref{fig:ellipse}, we show the expected 95\% confidence
level errors on the slope parameter $\beta_s$ on the halo occupation
number and the overall bias factor describing the large angular scale
clustering.  To recover the occupation number in detail, clustering
measurements at smaller angular scales are required to probe the
1-halo part and this is not possible with, for example, Planck high
frequency data alone.  As shown in Fig.~5, the required measurements
can be easily achieved with Herschel-SPIRE since Planck channels do
not have the adequate resolution. Furthermore, the combination of
Planck and Herschel over 2500 deg.$^2$ allows estimates of the
occupation numbers and the bias factor at the level of a few percent
at the 95\% confidence level even after accounting for the uncertain
redshift distribution of sources below the point source detection
level. For smaller area surveys down to the same depth, there is a
general degradation on parameter determination with the factor
$\sqrt{f_{\rm sky}}$.

In practice, once anisotropy measurements become available, clustering
analyses can be improved by combining unresolved fluctuations with
information from the clustering of resolved sources, number counts,
and luminosity functions.  The mechanisms to carry out such studies
already exist (See example involving 3.6$\mu$m Spitzer data in
Sullivan et al. 2007), but what is now clearly needed is a survey of
required area and sensitivity. Here we have shown that a survey of
order 10$^3$ deg.$^2$ provides maximal information on the clustering
of unresolved fluctuations.  

Finally, it may also be possible to use these anisotropy maps for a
weak lensing analysis in the same manner CMB maps are now proposed for
lensing studies given the large magnification bias at far-IR
wavelengths (e.g., Blain 1998).  The same maps can also be extended to
cross-correlate with Planck temperature anisotropy data at low
frequencies and at large angular scales to detect the integrated
Sachs-Wolfe (ISW) effect and at small angular scales to detect the CMB
lensing-far IR source cross-correlation.  The latter lensing-source
cross-correlation has been detected with WMAP and NVSS radio survey at
the 2$\sigma$ confidence level (Smith et al. 2007), but when Planck
data are combined with a wide-survey of Herschel of order 1000
deg$^2$, the cross-correlation can be detected at $>$ 20 $\sigma$
confidence level (Song et al. 2003). Once a better understanding
1/f-noise etc become available, it may be useful to returns to these
topics to exploit the full information content of Herschel.

\section{Summary}

Below the point source detection limit in upcoming far-IR surveys with
Planck and Herschel-SPIRE, correlations in the large-scale structure
will lead to clustered anisotropies in the unresolved component of the
far-infrared background (FIRB). The angular power spectrum of the FIRB
anisotropies could be measured in these surveys and will be one of the
few limited avenues to study some, though limited, information on the
faint sources that dominate the background light at these wavelengths.

To study the statistical properties of these anisotropies, the confusion
from foreground Galactic dust emission needs to be reduced even in the
``cleanest'' regions of the sky.  The multi-frequency coverage of
Planck and Herschel-SPIRE instrument allows the foreground dust to be
partly separated from the extragalactic background composed of dusty
starforming galaxies as well as faint normal galaxies.  The separation
improves for fields with sizes greater than a few hundred square
degrees and when combined with Planck data.  Here, we have shown that
an area of about $\sim$ 400 degrees$^2$ observed for about 1000 hours
with Herschel-SPIRE and complemented by Planck provides maximal
information on the anisotropy power spectrum.

Assuming such a survey will be conducted, we have discussed the
scientific studies that can be done with measurements of the
unresolved FIRB anisotropies including a few percent accurate
determination of the large scale bias and the small-scale halo
occupation distribution of FIRB sources with fluxes below the
point-source detection level. In practice, in addition to the
clustering spectrum of unresolved anisotropies, measurements such as
the luminosity function and the correlation function of resolved
sources as well as their redshift distribution must be modeled within
the same framework. Here, we have provided a detailed outline of such
a strategy based on the conditional luminosity functions associated
with the halo approach to large-scale structure.

\vspace{0.5cm}

{\it Acknowledgments:} 

We thank SPIRE SAG-1 and Open-Time Key Projects groups for useful
discussions. We thank G. Lagache and her collaborators for making
results of their source modeling publicly available in electronic
form.  This work was supported at UC Irvine by a McCue Fellowship (to
AA), NSF CAREER AST-0645427, and NASA support for science studies with
Herschel-SPIRE Guaranteed Time Observations at UC Irvine with JPL
Contract 1295096).


\begin{thebibliography}{99}
\frenchspacing

\bibitem[Berlind et al.<2003>]{Berlind:03}
Berlind, A.~A.,  Weinberg, D. H., Benson, A. J. et al. 2003, ApJ, 593, 1

\bibitem[Blain(1998)]{Bla98}
Blain, A. W. 1998, MNRAS, 295, 92

\bibitem[Blain et al.<2002>]{Bla02}
  Blain, A. W., Smail, I., Ivison, R. J., Kneib, J.-P. \& Frayer, D. T., 2002, Phys. Rept. 369, 111

\bibitem[Blain et al.<2004>]{Bla04}
  Blain, A. W., Chapman, S. S., Smail, I. \& Ivison, R. J. 2004, ApJ, 611, 725

\bibitem[Cooray \& Milosavljevi\'c<2005>]{CooMil:05}
Cooray, A., \& Milosavljevi\'c, M.\ 2005, ApJ, 627, L89

\bibitem[Cooray <2005>]{Cooray:05}
Cooray, A. 2005, MNRAS, 365, 842

\bibitem[Cooray \& Sheth(2002)]{CooShe02}
  Cooray, A. \& Sheth, R. 2002, Physics Reports, 372, 1 (astro-ph/0206508)

\bibitem[Crawford (2007)]{Cra07}
  Crawford, T. 2007, astro-ph/0702608

\bibitem[Dwek et al. (1998)]{dwek98} Dwek, E., et al. 1998, ApJ, 508, 106

\bibitem[Griffin et al. (2006)]{Gri06}
  Griffin, M. et al. 2006, in   ``Studying Galaxy Evolution with Spitzer and Herschel'', Crete (astro-ph/0609830)

\bibitem[Finkbeiner et al. (1999)]{Finetal99}
  Finkbeiner, D.P., Davis, M. \& Schlegel, D.J. 1999,
  ApJ,  524, 867

\bibitem[Fixsen et al. (1998)]{Fixsenetal98}
   Fixsen, D.J., Dwek, E., Mather, J.C., Bennett, C.L. \& Shafer, R.A.  1998,	
  ApJ, 508, 123	

\bibitem[Haiman \& Knox (2000)]{HaiKno00}
  Haiman, Z. \& Knox, L. 2000, ApJ, 530, 124

\bibitem[Hauser \& Dwek (2001)]{HawDwe01}
  Hauser, M. \& Dwek, E. 2001, ARAA, 39, 249

\bibitem[Knox et al. (2001)]{Knoetal01}
  Knox, L., Cooray, A., Eisenstein, D. \& Haiman, Z. 2001, ApJ, 550, 7

\bibitem[Kravtsov et al.<2004>]{Kravtsov:04}
Kravtsov, A.~V., Berlind, A.~A., Wechsler, R.~H., Klypin, A.~A., Gottl{\" o}ber, S., Allgood, B., \& Primack, J.~R.\ 2004, ApJ, 609, 35

\bibitem[Lagache et al. (2003)]{Lagetal02}
  Lagache, G., Dole, H. \& Puget, J.-L. 2003, MNRAS, 338, 555

\bibitem[Lagache et al. (2005)]{Lag05}
Lagache, G., Puget, J.-L., Dole, H. 2005, ARA\&A, 43, 727

\bibitem[Limber]{Lim54}
Limber, D. 1954, ApJ, 119, 655

\bibitem[Lin \& Mohr (2004)]{LinMoh04}
Lin, Y.-T. \& Mohr, J. J. 2004, ApJ, 617, L879


\bibitem[Negrello et al. (2007)]{neg07}
  Negrello, M. et al. 2007, astro-ph/0703210

\bibitem[Puget et al. (1996)]{puget96} Puget, J. L. et al. 1996, A\&A, 308,
L5

\bibitem[Schlegel et al. (1998)]{SFD98}
  Schlegel, D.J., Finkbeiner, D.P. \& Davis, M. 1998,
  ApJ,  500, 525

\bibitem[Scott \& White (1999)]{ScoWhi99}
  Scott, D. \& White, M. 1999, A\&A, 346, 1

\bibitem[Smith et al(2007)]{Smi07}
  Smith, K. M., Zahn, O. \& Dor\'e, O. 2007, arXiv:0705.3980

\bibitem[Song et al(2003)]{Song03}
  Song, Y.-S., Cooray, A., Knox, L. \& Zaldarriaga, M. 2003, ApJ, 590, 664

\bibitem[Sullivan et al. (2007)]{Sul07}
  Sullivan, I. et al. 2007, ApJ, 657, 37 (astro-ph/0609451)

\bibitem[Tegmark et al. (2003)]{Tegetal03}
  Tegmark, M., de Oliveira-Costa, A. \& Hamilton, A. 2003,
  Phys.\ Rev.\ D, 68, 123523

\bibitem[Tegmark et al. (1997)]{Teg97}
  Tegmark, M., Taylor, A. \& Heavens, A. 1997, ApJ, 480, 22

\bibitem[Vale \& Ostriker<2004>]{Vale:04}
Vale, A., \& Ostriker, J.~P.\ 2004, MNRAS, 353, 189


\bibitem[Yang, Mo, \& van den Bosch<2003>]{Yang:03b}
Yang, X., Mo, H.~J., \& van den Bosch, F.~C.\ 2003, MNRAS, 339, 1057

\bibitem[Yang et al.<2005>]{Yang:05}
Yang, X., Mo, H.~J., Jing, Y.~P., \& van den Bosch, F.~C.\ 2005, MNRAS, 358, 217



\end{thebibliography}
\end{document}